\documentclass[12pt]{article}

\usepackage{euscript,amsmath,amssymb,epsfig,latexsym,amsfonts}

\usepackage{graphicx}
\usepackage[english]{babel}

\newcommand{\be}{\begin{equation}}
\newcommand{\ee}{\end{equation}}
\newcommand{\bea}{\begin{eqnarray}}
\newcommand{\eea}{\end{eqnarray}}

\begin{document}
\title{\bf Self-adjoint approximations of \\  degenerate Schr\"{o}dinger operator}
\author{V.Zh. Sakbaev and I.V. Volovich\footnote {\small{\it volovich@mi.ras.ru}}\\~\\
\small{\it Steklov Mathematical Institute, RAS, Moscow }\\
}
\date {}
\maketitle

\abstract{The problem of construction  a quantum mechanical  evolution for the Schr\"{o}dinger equation with a degenerate Hamiltonian
which is a symmetric operator that does not have self-adjoint extensions
is considered. Self-adjoint regularization of the Hamiltonian does not lead to a preserving probability limiting evolution for vectors from the Hilbert space but
it is used to construct a limiting evolution of states on a  $C^*$-algebra
of compact operators and on an abelian subalgebra of operators in the Hilbert space. The limiting evolution of the states on the abelian algebra can be presented by the Kraus decomposition with two terms. Both of this terms are corresponded to the unitary and shift components of Wold's decomposition of isometric semigroup generated by the degenerate  Hamiltonian.   Properties of the limiting evolution of the states on the C*-algebras     are investigated and it is shown that pure states could evolve into mixed states.

}

\section{Introduction}
Degenerate elliptic and parabolic equations are widely studied,
see for example \cite{Fichera ,Oleinik-Radkevich ,Freidlin-Ventcel'}.
Universal boundary conditions for various types of PDE were considered in \cite{Sakb-Vol}.

It was shown \cite{Oleinik-Radkevich ,Freidlin-Ventcel'} that the solvability of the Cauchy problem for the degenerate differential equation with smooth coefficients in the whole space holds but the degenerate differential equation without smoothness of coefficients can have no solution.  We will study an  example of the simplest first order differential operator on the semi-line. Note that the boundedness of the domain in Euclidian coordinate space can arise as the result of singularity of coefficients in the complement of this domain (see \cite{Sakb-Smol}).

In this paper we concern with the degenerate Schr\"{o}dinger equation, see \cite{STMF, SFPM, SR11} of the form
\be \label {schr}
i\frac{\partial u}{\partial t}={\bf H}u
\ee
where $-{\bf H}$ is a second order differential operator with nonnegative
characteristic form in a region $G\subset\mathbb{R}^n$. It is assumed that under some boundary conditions ${\bf H}$ defines a symmetric operator in
the Hilbert space $H=L^2(G)$.  We are interested in the case when ${\bf H}$ does not have self-adjoint extensions.

The initial-boundary value problem for the Schrodinger equation \ref{schr}
does not necessary has a solution. In such a case we will use
an elliptic regularization and proceed as follows. Consider a family
of operators ${\bf H}_{\epsilon}$ depending on the parameter $\epsilon >0$,
\be\label{epsil}
{\bf H}_{\epsilon}=-\epsilon\Delta +{\bf H}
\ee
where $\Delta$ is the Laplace operator. Suppose that ${\bf H}_{\epsilon}$
admits a  self-adjoint extension (we denote it by the same letter).
Let $u_{\epsilon}(t)=U_{\epsilon}(t)\varphi$ be a solution of the
Cauchy problem
\be\label{cauchy}
i\frac{\partial u_{\epsilon}}{\partial t}={\bf H}_{\epsilon}u_{\epsilon}
\ee
\be
u_{\epsilon}(0)=\varphi, \,\,\,\,\varphi\in L^2(G)
\ee
Here $U_{\epsilon}(t)=e^{-it{\bf H}_{\epsilon}}$ is the unitary evolution operator.

Shchrodinger equation with small paremeter arise in the study of limit behavior of a class of a such quantum systems as electrons and holes in the nonhomogenious semiconducte materials in the case of large value of effective mass of a particles in one part of materials and small value of effective mass in another  (see \cite{Gadella, GS, SR11}).

The limit of $u_{\epsilon}(t), \epsilon\to 0$ in the Hilbert space
might be not exist however we will show that in some cases
there exists the limit
\be\label{qlim}
\sigma_{\varphi}^t(A)=\lim_{\epsilon\to 0}\langle u_{\epsilon}(t),Au_{\epsilon}(t)\rangle=
\lim_{\epsilon\to 0}\langle \varphi,U_{\epsilon}^*(t)AU_{\epsilon}(t)\varphi\rangle
\ee
for any $\varphi\in H, t\geq 0$ where $A$
is an operator from a subalgebra  of
the algebra $B(H)$ of bounded operators in $ H$.

We will show that $\sigma_{\varphi}^t$ defines a state (a linear positive functional of norm 1) on a $C^*$-algebra and therefore it can be interpreted   as quantum mechanical evolution
generated by the Schrodinger operator ${\bf H}$.

Degenerate Hamiltonians arise in  different problems.

1. The theory of semi-classical limit is the theory of the passage to the limit in the family of second order differential operators with the small paremeter at the second derivatives. The vanishing of the coefficients with second derivative in the case is the degeneration of Hamiltonians; the limit operator is the first order differential operator which can be symmetric but not self-adjoint operator.

2. Similar effects has the family of Hamiltonians with large mass parameter. In the semiconductors theory (see \cite{Gadella, GS}) the mass of a system is the nonnegative function on the coordinate spece wich values is the constant in some domain of coordinate space and is the large parameterin the complement of this domain. Then the limit Hamiltonian is the degenered symmetric but not self-adjoint differential operator of second order in the domain and of first order in its complement (see \cite{SFPM, SR11, Sakb-Smol}).

3. Symmetric differential operators arise in the theory of conductance in the geometrical graphs in nano-semiconductors theory (see \cite{Sakb-Smol, Shafarevich}). The degeneration of the second orger coefficients of diffferential operator on a graph can be the reason of the absence of self-adjoint extensions of this operators.

4. The both of degeneration and nonsmoothness of coefficients of a first order transport equation on a line is the reason of nonexistence or nonuniqueness of its solutions. Hence it is the reason of luck of self-adjointness of the first order operator (see \cite{Freidlin-Ventcel', DPL}).

5. Positive-operator valued measures (POVM) are used in the quantum measurement theory.
Neurmak's dilation theorem states that a POVM can be lifted to a projection-valued measure. It is used in theory of open quantum systems (see \cite{OV,ALV,TV,AVK,VK}).

In the next section we study a simple model of quantum dynamics on the half-line with the symmetric Hamiltonian which does not admit a self-adjoint extension.

 \section{Schrodinger equation on the half-line}

 In quantum mechanics, according to von Neumann's axioms, observables correspond to self-adjoint operators.
 However, Hermitian (symmetric) operators which don't admit
 self-adjoin extensions are also discussed  as  possible observables (see for example \cite{OV}).

It is well known that the momentum operator on a half-line is not self-adjoint. Here we consider the momentum operator on the half-line $\mathbb{R}_+=\{x\in \mathbb{R}:x>0\}$ as an example of a degenerate  Hamiltonian for the Schr\"{o}dinger equation and try to define an appropriate quantum dynamics. Let the Hamiltonian
will have the form
\be\label{dif}
{\bf H}=-b\hat{p}=
ib\frac{\partial}{\partial x}\,,\,\,\,b\in \mathbb{R}
\ee
The differential operator $H$ defines a symmetric operator
(denoted by the same letter) in the Hilbert space $ H=L_2(\mathbb{R}_+)$ with the domain
$W_{2,0}^1(\mathbb{R}_+)$ which is a completion of the space $C_0^{\infty}(\mathbb{R}_+)$ with respect to the norm of the Sobolev space  $W_{2}^1(\mathbb{R}_+)$, in particular $u(+0)=0$
if $u\in W_{2,0}^1(\mathbb{R}_+)$. The adjoint operator $H^*$
is defined on the domain $W_{2}^1(\mathbb{R}_+)$ by the relation (\ref{dif}). The deficiency
index of the operator  ${\bf H}$ is $(n_+,n_-)=(1,0)$ if $b<0$ and
$(0,1)$ if $b>0$, so the operator $H$ doesn't have self-adjoint extensions.

The Schrodinger equation with the Hamiltonian (\ref{dif})
reads
\begin{equation}\label{S-Schr-5}
\frac{\partial u}{\partial t}=
b\frac{\partial u}{\partial x},\,\,t>0,\,x>0,
\end{equation}
with the initial-boundary value data \be\label{cauch}
u(+0,x)=\varphi(x)\,,\,\,x>0, \ee \be\label{bound12}
u(t,+0)=0\,,\,\,t>0. \ee The problem (\ref{S-Schr-5}) -
(\ref{bound12}) does not have a solution in the case $b>0$ since
the  solution $u(t,x)=\varphi(bt+x)$ of (\ref{S-Schr-5}) -
(\ref{cauch}) does not satisfy the boundary condition
(\ref{bound12}) if $b\neq 0,\,\varphi\neq 0$.

But in the case $b< 0$ the problem (\ref{S-Schr-5}) -  (\ref{bound12}) with the initial data $\varphi \in D({\bf H})$ has a unique solution  which is given by the formula
$u(t,x)=\varphi(bt+x)\,$ if $\, bt+x>0;\quad u(t,x)=0\,$ if $\, bt+x\leq 0;\ (t,x)\in \mathbb{R}_+\times \mathbb{R}_+.$ The uniqueness of this solution is the consequence of the symmetricity of Hamiltonian ${\bf H}$.

To obtain a unified approach to this two cases we consider an
approximate Hamiltonian \be \label{regul}
{\bf H}_{\epsilon}=-\epsilon\frac{\partial^2}{\partial x^2}
+ib\frac{\partial}{\partial x} \ee where $\epsilon >0$. It defines
a self-adjoint operator with the domain $D({\bf H}_{\epsilon})=\{u\in
W_{2}^2(\mathbb{R}_+):u(+0)=0\}$.

The corresponding Schr\"{o}dinger equation reads
\begin{equation}\label{S-Schr-2}
i\frac{\partial u_{\epsilon}}{\partial t}=-\epsilon \frac{\partial^2 u_{\epsilon}}{\partial x^2}
+ib\frac{\partial u_{\epsilon}}{\partial x},\,\,t>0,\,x>0,
\end{equation}
with the initial-boundary data
\begin{equation}\label{Cau-2}
u_{\epsilon}(+0,x)=\varphi(x),\,\,\, x>0,\, \varphi\in  H
\end{equation}
\be\label{Cau-2b}
u_{\epsilon}(t,+0)=0,\,\,t>0.
\ee

A classical solution of the problem (\ref{S-Schr-2}) - (\ref{Cau-2b}) on the interval $I=(0,T),\, T>0$ is a function $v(t)\in C(I,C^2(\mathbb{R}_+)\cap D(H_{\epsilon}))\cap C^1(\bar{I},C(\mathbb{R}_+))
$ which satisfies Eq.(\ref{S-Schr-2})
and the boundary conditions (\ref{cauch}) - (\ref{bound12}).

A generalized solution of the problem (\ref{S-Schr-2}) - (\ref{Cau-2b}) on the interval $I=(0,T),\, T>0$ is a function $v(t)\in C(I,H)
$ which satisfies the integral equality
$$
(v(t),\psi )=(\varphi ,\psi )+i\int\limits_0^t(v(s),{\bf H}_{\epsilon }\psi )ds ,\, t\in R_+,
$$
for arbitrary element $\psi \in D({\bf H}_{\epsilon})$.

 The uniqueness of solution of the Cauchy problem  (\ref{S-Schr-2}) - (\ref{Cau-2b}) is the consequence of the self-adjointness of the Hamiltonian ${\bf H}_{\epsilon}$.

We will show that the asymptotic behavior of the solution   $u_{\epsilon}(t,x)$ of the problem (\ref{S-Schr-2}) - (\ref{Cau-2b}) as $\epsilon\to 0$ has the form
\begin{equation}\label{as-3}
u_{\epsilon}(t,x)\sim
\Psi_{\epsilon}(t,x)=\varphi(bt+x)-e^{i\frac{b}{\epsilon}x}\varphi(bt-x)
\end{equation}
Note that
\begin{equation}\label{as-4}
 \Psi_{\epsilon}(t,0)=0,\,\,t>0
\end{equation}

Firstly we obtaine the representation of the solution of the problem  (\ref{S-Schr-2}) - (\ref{Cau-2b}) for the sufficiently smooth initaial data $\varphi$ and after that we extend this result to arbitrary initial data.

{\bf Lemma 1}. Let $b>0$ and $\varphi\in W^3_{2,0}(\mathbb{R}_+)  $.
Then the solution $u_{\epsilon}$ of the problem (\ref{S-Schr-2}) - (\ref{Cau-2b}) can be represented in the form
\begin{equation}\label{sol-4}
u_{\epsilon}(t,x)
=\frac{e^{-i\pi/4}}{\sqrt{4\pi \epsilon t}}\int_0^{\infty} [e^{\frac{i}{4\epsilon t}(x-y+bt)^2}-
e^{\frac{i}{4\epsilon t}(x+y-bt)^2}e^{\frac{ib}{\epsilon}x}]
\varphi(y)dy.
\end{equation}

{\bf Proof}. Let $u_{\epsilon}$ is a solution of Eq.(\ref{S-Schr-2}). Then the function
\begin{equation}\label{deff}
 v(t,x)=e^{-\frac{ib^2}{4\epsilon}t-\frac{ib}{2\epsilon}x}
u_{\epsilon}(t,x)
\end{equation}
satisfies the following Schr\"{o}dinger equation for the free particle
\begin{equation}\label{Schr-9}
i\frac{\partial v}{\partial t}=-\epsilon \frac{\partial^2 v}{\partial x^2}\,,\,\,t>0\,,\,\,x>0.
\end{equation}
Equivalent problem for Eq.(\ref{Schr-9})
has the following initial-boundary value conditions:
\begin{equation}\label{Cau-5}
v(+0,x)=e^{-\frac{ib}{2\epsilon}x}\varphi(x),\,\,\, x\in \mathbb{R}_+,
\end{equation}
\begin{equation}\label{Cau-6}
v(t,+0)=0,\,\,\, t\in \mathbb{R}_+
\end{equation}
 It is known that the solution of the problem (\ref{Schr-9}), (\ref{Cau-5}), (\ref{Cau-6})
 has the form
\begin{equation}\label{sol-3}
v(t,x)=\frac{e^{-i\pi/4}}{\sqrt{4\pi \epsilon t}}\int_0^{\infty} [e^{\frac{i}{4\epsilon t}(x-y)^2}-
e^{\frac{i}{4\epsilon t}(x+y)^2}]e^{\frac{ib}{2\epsilon}y}\varphi(y)dy
\end{equation}
Therefore the solution of the problem (\ref{S-Schr-2}) - (\ref{Cau-2b})
will have the form:
\begin{equation}\label{sol-4a}
u_{\epsilon}(t,x)=\frac{e^{-i\pi/4}}{\sqrt{4\pi \epsilon t}}e^{\frac{ib^2}{4\epsilon}t+\frac{ib}{2\epsilon}x}\int_0^{\infty} [e^{\frac{i}{4\epsilon t}(x-y)^2}-
e^{\frac{i}{4\epsilon t}(x+y)^2}]e^{\frac{ib}{2\epsilon}y}
\varphi(y)dy
\end{equation}
$$
=\frac{e^{-i\pi/4}}{\sqrt{4\pi \epsilon t}}\int_0^{\infty} [e^{\frac{i}{4\epsilon t}(x-y+bt)^2}-
e^{\frac{i}{4\epsilon t}(x+y-bt)^2}e^{\frac{ib}{\epsilon}x}]
\varphi(y)dy.
$$

Note that under the assumption $\varphi \in W^3_{2,0}(R_+)$ we have $$u_{\epsilon}\in C(\mathbb{R}_+,W^3_{2,0}(\mathbb{R}_+))\bigcap C^1(\mathbb{R}_+,W^1_{2,0}(\mathbb{R}_+)).$$ Hence the function $u_{\epsilon}$ is the classical solution for the problem (\ref{S-Schr-2}) - (\ref{Cau-2b}).
Since the space $W^3_{2,0}(\mathbb{R}_+)$ is dense in the space $ H=L_2(\mathbb{R}_+)$ and since the map ${\bf U}_{\epsilon }(t)$
$$
W^3_{2,0}(\mathbb{R}_+)\ni \varphi (\cdot) \ \to \ u_{\epsilon }(t,\cdot )
$$
for any $t\geq 0$ preserves the $L_2$-norm: $\|u_{\epsilon}
(t,\cdot)\|_H=\| \varphi \|_H$ then the family of
maps ${\bf U}_{\epsilon }(t),\, t\geq 0,$ can be redefined by the
continuity onto the unitary semigroup
\begin{equation}\label{grupp}
{\bf U}_{\epsilon }(t),\, t\geq 0,
\end{equation}
in the space
$ H$. Then for any $\varphi \in H$ the equality  (\ref{sol-4a}) gives the function ${\bf
U}_{\epsilon}(t)\varphi $
and this function is the
generalised solution of the problem (\ref{S-Schr-2}) -
(\ref{Cau-2b}).

{\bf Theorem 1}. Let  $b>0$ and $\varphi\in  H$. Then for the solution $u_{\epsilon }$ (\ref{sol-4}) of the problem (\ref{S-Schr-2}) - (\ref{Cau-2b})
the following asymptotic expansion holds as $\epsilon\to 0$
$$
u_{\epsilon}=v_{\epsilon}+r_{\epsilon },\ \epsilon \in (0,1),
$$
where
\begin{equation}\label{asol'-4}
v_{\epsilon }(t,x)=\varphi(x+bt)-\theta (bt-x)\varphi(bt-x)e^{\frac{ib}{\epsilon}x},\ (t,x)\in \mathbb{R}_+\times \mathbb{R}_+,
\end{equation}
and
\be\label{est}
\lim\limits_{\epsilon \to +0}(\sup\limits_{t\in [0,T]}\| r_{\epsilon }(t,\cdot )\|_{ H})=0
\ee
for any $T>0$.

{\bf Proof}.

The proof of the theorem 1 consists of a several lemmas.

{\bf Lemma 2.} Let  $b>0$ and $\varphi\in  H$. Then for any $T>0$
\be
\lim\limits_{\epsilon \to +0}(\sup\limits_{t\in [0,T]}\| \varphi(x+bt)- \frac{e^{-i\pi/4}}{\sqrt{4\pi \epsilon t}}\int_0^{\infty} e^{\frac{i}{4\epsilon t}(x-y+bt)^2}
\varphi(y)dy\|_{ H})=0.
\ee

Let $v_0\in L_2(\mathbb{R})$ be the function which is defined by the conditions:  $v_0|_{\mathbb{R}_-}=0$ and $v_0|_{\mathbb{R}_+}=\varphi$. Then for any $t>0$ the following function $v_{\epsilon}(t,\cdot ):\ mathbb{R}\to \mathbb{C}$ is defined by the equality
\begin{equation}
\label{padenie}
v_{\epsilon }(t,x)=\frac{e^{i\pi/4}}{\sqrt{4\pi \epsilon t}}\int_{-\infty }^{\infty} e^{-\frac{i}{4\epsilon t}(x-y+bt)^2}v_0(y)dy,\quad x\in R,\, t>0.
\end{equation}
Hence Foirier image $\hat v(t,\xi )={\cal F}(v(t,\cdot ))(\xi )$ satisfies the equality
$$
\hat v _{\epsilon }(t,\xi )=e^{-2i\epsilon t\xi^2-ibt\xi}\hat v_0(\xi).
$$
Since $u_0\in L_2(\mathbb{R}_+)$, then $v_0\in L_2(\mathbb{R})$ and $\hat v_0\in L_2(\mathbb{R})$.
Therefore for any $\sigma >0$ there is a number $L>0$ such that $\|\hat u_0-{\bf P}_L\hat u_0\|_{L_2(\mathbb{R})}<{{\sigma }\over 4}$ where ${\bf P}_{L}$ is the operator of multiplication onto the indicator function $\chi _{[-L,L]}$ of the segment $[-L,L]$. Since the functions $\{ i\epsilon t\xi ^2,\, \epsilon >0$, tends to zero as $\epsilon \to 0$ uniformly on any rectangular $[0,T]\times [-L,L]$, then there is the number $\epsilon _0>0$ such that the estimate $\sup\limits_{t\in [0,T]}\int\limits_{-L}^L|(e^{i\epsilon t\xi^2}-1)e^{ibt\xi}\hat u_0(\xi )|^2d\xi \leq {{{\sigma }^2}\over 2}$ holds for any $\epsilon \in (0,\epsilon _0)$.

Thus there is the number $\epsilon _0>0$ such that the estimates $\sup\limits_{t\in [0,T]}\| \hat v_{\epsilon }(t,\xi )-e^{ibt\xi}\hat v_0(\xi)\|_{L_2(\mathbb{R})}^2=\sup\limits_{t\in [0,T]}(\int\limits_{-\infty }^{-L}+\int\limits_{-L }^{L}+\int\limits_L^{\infty })|(e^{i\epsilon t\xi^2}-1)e^{ibt\xi}\hat u_0(\xi )|^2d\xi \leq \|2(u_0-{\bf P}_Lu_0)\|_{L_2(R)}^2+\sup\limits_{t\in [0,T]}\int\limits_{-L}^L|(e^{i\epsilon t\xi^2}-1)e^{ibt\xi}\hat u_0(\xi )|^2d\xi \leq \sigma ^2$ hold for any $\epsilon \in (0,\epsilon _0)$.

Since $\sigma >0$ and $T>0$ are arbitrary numbers then the sequence of functions $
\hat v _{\epsilon },\, \epsilon \to 0,$ converges in the space $L_2(\mathbb{R})$ to the function $\hat w(t,\xi )=e^{ibt\xi}\hat v_0(\xi)$ uniformly on the seqment $[0,T]$ with respect to parameter  $t$. According to the unitarity of Fourier transformation the sequence of functions   $ v _{\epsilon },\, \epsilon \to 0,$ converges in the space  $L_2(\mathbb{R})$ to the function $w(t,x)=v_0(x+bt),\, x\in \mathbb{R},$ uniformly on the segment $[0,T]$ with respect to the parameter  $t$. Since the functions $u_{\epsilon }$ are the restrictions of the functions $v_{\epsilon }$ on the quadrant $K=\{ t>0,\, x>0\}$ then the sequence of functions $ u _{\epsilon },\, \epsilon \to 0,$ converges in the space $L_2(\mathbb{R}_+)$ to the function $u(t,x)=v_0(x+bt)|_{K}$  uniformly on the segment $[0,T]$ with respect to parameter  $t$ (note that if $b>0$ then $v_0(x+bt)|_{K}=\varphi (x+bt)|_{K}$).
Lemma 2 is proved.

{\bf Lemma 3.} Let  $b>0$ and $\varphi\in H$. Then for any $T>0$
\be
\lim\limits_{\epsilon \to +0}(\sup\limits_{t\in [0,T]}\| \theta (bt-x)\varphi(bt-x)- \frac{e^{-i\pi/4}}{\sqrt{4\pi \epsilon t}}\int_0^{\infty} e^{\frac{i}{4\epsilon t}(x+y-bt)^2}
\varphi(y)dy\|_{ H})=0.
\ee

To investigate the limit behavior of the second term in the expression  (\ref{sol-4}) we should introduced the following sequence of functions $\{\psi _{\epsilon }:\ (0,+\infty )\times \mathbb{R}\, \to \, \mathbb{R},\ \epsilon \in (0,1)\}$, where for any  $t>0$ the function $\psi _{\epsilon}(t,cdot )$ is defined on the line $\mathbb{R}$ by the equality
\begin{equation}
\label{otrazhenie}
\psi _{\epsilon }(t,x)=\frac{e^{i\pi/4}}{\sqrt{4\pi \epsilon t}}\int_0^{\infty}
e^{-\frac{i}{4\epsilon t}(x+y-bt)^2}
v_0(y)dy,\quad x\in \mathbb{R},\, t>0.
\end{equation}
By means  of the convergence of the sequence (\ref{padenie}) we can prove that the sequence of functions $ \psi _{\epsilon },\, \epsilon \to 0,$ converges in the space  $L_2(\mathbb{R})$ to the function $\psi (t,x)=v_0(bt-x)$  uniformly on the segment $[0,T]$ with respect to parameter  $t$. Hence the sequence of its restrictions on the quadrant $K$ converges to the restriction on the quadrant $K$ of a function  $\psi (t,x)=v_0(bt-x)$ (note that the support of the last function is ste subset of the angle $x>0,\, bt-x>0$).

Hence the equality
$$
\lim\limits_{\epsilon \to 0}\sup\limits_{t\in [0,T]}\| \frac{e^{i\pi/4}}{\sqrt{4\pi \epsilon t}}e^{-i{b\over {\epsilon }}x}\int_0^{\infty}
e^{-\frac{i}{4\epsilon t}(x+y-bt)^2}
u_0(y)dy - e^{-i{b\over {\epsilon }}x}\psi (t,x)\|_{L_2(\mathbb{R}_+)}=0
$$
holds for any $T>0$.

Therefore the equalities (\ref{asol'-4})-(\ref{est}) hold as the consequence of Lemmas 2 and 3.

\bigskip

{\bf The convergence of the sequence  of regularized unitary groups
$\{ {\bf U}_{\epsilon}\}$ acting in the space $H$}

Let the operator $\bf H$ in the space $H$ is defined by the
relation (\ref{dif}) and the sequence of regularized Hamiltonians
${\bf H}_{\epsilon },\ \epsilon >0, \, \epsilon \to 0$ is given by
the expression (\ref{regul}).

Then according to the theorem 1 we obtain the following results
about the behavior of the sequence $\{{\bf U}_{\epsilon }\}$ as
$\epsilon \to 0$, where ${\bf U}_{\epsilon }(t)=e^{-it{\bf
H}_{\epsilon }},\, t\in \mathbb{R}$.

{\bf Proposition 1} (\cite{SR11}). {\it Let $\bf H$ is maximal symmetric operator
in the space $H$ with deficient indexes $(n_-,n_+)$. Then

1. If $n_+=0$ then the operator $-i{\bf H}$ is the generator of
isometric semigroup ${\bf U}_{\bf L}(t)=e^{-it{\bf L}},\, t\geq
0,$ and the operator $-i{\bf H}^*$ is the generator of contractive
semigroup ${\bf U}_{{\bf L}^*}(t)=e^{-it{\bf H}*},\, t\leq 0$.

2. If $n_-=0$ then the operator $-i{\bf H}$ is the generator of
isometric semigroup ${\bf U}_{\bf L}(t)=e^{-it{\bf L}},\, t\leq
0,$ and the operator $-i{\bf H}^*$ is the generator of contractive
semigroup ${\bf U}_{{\bf L}^*}(t)=e^{-it{\bf H}*},\, t\geq 0,$.
}

For any $b\in R$ we define the operator-valued function ${\bf
V}_b$ by the following conditions:

1. If $b\leq 0$ then ${\bf V}_b(t)=e^{-it{\bf H}},\, t\geq 0$ and
${\bf V}_b(t)=e^{-it{\bf H}^*},\, t\leq 0$.

2. If $b\geq 0$ then ${\bf V}_b(t)=e^{-it{\bf H}},\, t\leq 0$ and
${\bf V}_b(t)=e^{-it{\bf H}^*},\, t\geq 0$.

{\bf Proposition 2} (\cite{SR11}). {\it Let the operator $\bf H$ in the space $H$
is defined by the relation (\ref{dif}) and the sequence of
regularized Hamiltonians ${\bf H}_{\epsilon },\ \epsilon >0, \,
\epsilon \to 0$ is given by the expression (\ref{regul}). Then the
sequence of unitary group $\{ {\bf U}_{\epsilon }\}$ converges in
the weak operator topology uniformly on any segment $[c,d]\in \mathbb{R}$
to the operator-valued function ${\bf V}_b$:
$$
\lim\limits_{\epsilon \to 0}\sup\limits_{t\in [c,d]}|(\phi , ({\bf
U}_{\epsilon }(t)-{\bf V}_b(t))\psi)|=0\quad \forall \ c,d\in \mathbb{R};\,
\phi,\psi \in H.
$$
Moreover, if $b\leq 0$ then
$$
\lim\limits_{\epsilon \to 0}\sup\limits_{t\in [c,d]}\|({\bf
U}_{\epsilon }(t)-{\bf V}_b(t))\psi)\|=0\quad \forall \ c,d\in
[0,+\infty );\, \psi \in H,
$$
or if $b\geq 0$ then
$$
\lim\limits_{\epsilon \to 0}\sup\limits_{t\in [c,d]}\|({\bf
U}_{\epsilon }(t)-{\bf V}_b(t))\psi)\|=0\quad \forall \ c,d\in
(-\infty ,0];\, \psi \in H.
$$
}

\section{Quantum dynamics on the half-line}

It is evident from Theorem 1 that the limit of the solution
$u_{\epsilon}$ as $\epsilon\to 0$ doesn't exist in ${\cal H}$ due to the oscillation factor $e^{ibx/\epsilon}$. However there exists
a weak limit and moreover we prove now that there exists the limit
(\ref{qlim}) for mean values of operators from  a $C^*$-subalgebra  of the $C^*$-algebra $ B( H)$ of bounded operators in $ H$.

In an algebraic approach to quantum mechanics \cite{BR} a physical
system is defined by its $C^*$-algebra ${\cal A}$ with identity.
Observables correspond to self-adjoint elements of ${\cal A}$.
The states are normalized positive linear functionals on ${\cal A}$. The time dynamics is given by a (semi)group of $\ast$-automorphism of ${\cal A}$.

\medskip

The quantum theory can be presented by the following data   $\{{\cal A}, S, \alpha^t   \}$, where ${\cal A}$ is the $C^*$-algebra (or von Neyman algebra), $S$ is the set of states on the algebra ${\cal A}$, $\alpha^t$ is the (semi)group of authomorphisms of the algebra ${\cal A}$. 
The (semi)group $T^t$ of dual authomorphisms of the set of states is defined by the formula

$$<T^t\rho,A>=
 <\rho, \alpha^t(A)>,\,\rho\in S, A\in{\cal A}.$$
In both cases the group property takes place: $\alpha^{t+\tau}=\alpha^{\tau}\alpha^t$ and  $T^{t+\tau}=T^tT^{\tau}$.

Any group $U_t$ of unitary operators in the Hilbert space $H$ generates the group of authomorphisms $\alpha ^t$ of the algebra ${\cal A}=B( H)$ of all bounded linear operators in the space $ H$ which is given by the formular $\alpha^t(A)=U^*_tAU_t$,\,\, $A\in {\cal A}$.

For investigation of our model we consider some different choises of a $C^*$-algebra $\cal A$ of observables: as the $C^*$-algebra $\cal A$ the following cases will be studied:

1. the algebra $B(H)$ of all bounded linear operators,

2. the algebra ${\cal A}_{comp}={\cal K}\bigcup \{ {\bf I}\}$
consists of the ring $\cal K$ of compact operators and the unity
operator,

3. the algebra ${\cal A}_{mult}$ of operators of multiplication on the measurable essentially bounded function which can be presented by the abelian algebra $L_{\infty}(R_+)$.

4. the algebra ${\cal A}_{sum}={\cal A}_{comp}\oplus {\cal A}_{mult}$ of operators which can be presented as the summ of two terms ${\bf A}_1\in {\cal A}_{comp}$ and  ${\bf A}_2\in {\cal A}_{mult}$.

We study the limit behavior as $\epsilon \to 0$ of the family of
the groups of automorphisms $\{\alpha ^t _{\epsilon},\, t\in R\}$
of $C^*$-algebra $\cal A$ and the family of the groups of
conjugate automorphisms $T^t_{\epsilon}$ of the set  of states
$\Sigma ({\cal A})$. For any $\epsilon
>0$ and $t\in R$ the map $\alpha ^t_{\epsilon}:\ {\cal A}\, \to \,
{\cal A}$ is defined by the equalities $ \alpha^t_{\epsilon}({\bf
A})=U_{\epsilon}^*(t){\bf A}U_{\epsilon}(t),\ {\bf A}\in {\cal A}.
$ The unitary groups ${\bf U}_{\epsilon }$ is generated by the
regularized Hamiltonian (\ref{regul}) according to the equality
\be\label{reggr} {\bf U}_{\epsilon }^t=e^{-it{\bf H}_{\epsilon}},\
t\in \mathbb{R}. \ee

Firstly  we consider the case ${\cal A}=B({\cal H}), \,\,
\alpha^t_{\epsilon}(A)=U_{\epsilon}^*(t)AU_{\epsilon}(t). $ The
regularized dynamic $T_{\epsilon}^t$ of  the set of states is
defined by the action of the map $T_{\epsilon}^t$ on any pure
state $\rho _{\varphi }$, where $\varphi$ is a unit vector from
the space ${\cal H}$:  $$<T_{\epsilon}^t\rho _{\varphi },A>=
<U_{\epsilon}(t)\varphi,AU_{\epsilon}(t)\varphi >.$$ Therefore the
group property for the regularized dynamics
holds: $\alpha^{t+\tau}_{\epsilon}=\alpha^{\tau}_{\epsilon}
\alpha^t_{\epsilon},\, \,
T^{t+\tau}_{\epsilon}=T^{\tau}_{\epsilon}T^t_{\epsilon}$.

Now we give the description of the set of quantum states on the algebra $B(H)$.  Let $B(H)$ be the Banach space of
linear bounded operators in the space $H$ and $B^{*}(H)$ is the
conjugate space. The set of quantum states $\Sigma (H)$ is the
set of linear continuous nonnegative normalized functionals on the
space $B(H)$: the set $\Sigma (H)$  is the intersection of the unique sphere with the positive cone of the space $(B(H))^*$ (see \cite{Takesaki}).

Let $\Sigma _n(H)$ be the set of all
quantum states which is continuous not only in the norm topology
but also in strong operator topology on the space $B(H)$. Let $T(H)$ be the Banach space of trace-class operators endowing with trace norm. According to the results by \cite{BR} an arbitrary
normal state $\rho \in \Sigma _{n}(H)$ can be uniquely
represented by some nonnegative trace-class operator $r\in T(H)$ with unique trace-norm.
Let $\Sigma _p(H)$ be the set of pure vector states. Any pure state can be uniquely represented by one
dimensional orthogonal projector. The pure state $\rho _{u}$ which is given by the orthogonal projector on the unique vector $u\in H$ defines the linear continuous functional $\rho
_{u}:\quad \langle \rho _{u},{\bf A}\rangle=(u,{\bf A}u)_{H},\, {\bf A}\in B(H)$, on the space $B(H)$. According to \cite{BR} the set $\Sigma _{p}(H)$ is the set of extreme points of the set $\Sigma _{n}(H)$.

The state $\rho \in \Sigma (H)$ is called singular state
iff the following equality holds $$\sup\limits_{\|u\|_H=1}\langle \rho
,{\bf P}_{u}\rangle =0$$ (where ${\bf P}_{u}$ is orthogonal
projector on the unit vector $u\in H$). The set of singular
states is denoted by $\Sigma _{s}(H)$. The
following analogue of the statement of Iosida-Hewitt theorem for measures takes place for the states.

{\bf Lemma 4.} (See \cite{Wils, PV}) {\it For any states  $\rho \in \Sigma (H)$ there
is the unique two functionals $r_{n}\in T(H),\, r_{b}\in
B^{*}(H)\backslash T(H)$ such that $\rho = r_{n}+r_{b}$.}

To investigate the asymptotics of the regularized trajectories $\rho _{u_{\epsilon }(t)},\, t\geq 0,$
in the space of quantum states we use the decomposition of the solution of regularized Cauchy problem (\ref{S-Schr-2}) - (\ref{Cau-2b})  in the form (see Lemma 1):
$$
u_{\epsilon }(t,x)=\phi _{\epsilon }(t,x)+\psi _{\epsilon }(t,x),
$$
where $\phi _{\epsilon }(t,x)=u_{\epsilon }(t,x)-\psi _{\epsilon }(t,x)$ and $\psi _{\epsilon }(t,x)=\theta (bt-x)\varphi (bt-x)e^{i{b\over {\epsilon}}x}$ (see theorem 1).
Then according to the theorem 1 we have that the following two statements hold:

{\bf Remark 1}. The sequence of functions $\{ \phi _{\epsilon }(t,x)\} $
converges in the space $C(R_+,H)$ uniformly on any segment $[0,T]$
to the limit function $v (t,x)=\varphi (x+bt)$ i.e. for any number
$T>0$ the equality holds
$$
\lim\limits_{\epsilon \to 0}\sup\limits_{t\in [0,T]}\|v (t)-\phi _{\epsilon }(t)\|_H=0;
$$
{\bf Remark 2.} the sequence of functions $\{ \psi _{\epsilon }(t,x)\} $
converges weakly in the space $C(R_+,H)$ to zero uniformly on any
segment $[0,T]$ i.e. for any function $g\in H$ and any number
$T>0$ the equality holds
$$
\lim\limits_{\epsilon \to 0}\sup\limits_{t\in [0,T]}|(g,\psi _{\epsilon }(t))_H|=0.
$$

{\bf Lemma 5.} For any operator ${\bf A}\in B(H)$ the equality $\lim\limits_{\epsilon \to 0}\langle \rho _{{u_{\epsilon}}},{\bf A}\rangle
=(v,{\bf A}v)+\lim\limits_{\epsilon \to 0}(\psi _{\epsilon},{\bf A}\psi _{\epsilon })$ holds.

In fact, for any $\epsilon >0$ the following equality:
\begin{equation}\label{lemma2}
(u_{\epsilon},{\bf A}u_{\epsilon })=(\phi_{\epsilon},{\bf A}\phi
_{\epsilon })+2{\rm Re}((\psi _{\epsilon},{\bf A}\phi _{\epsilon
}))+(\psi _{\epsilon},{\bf A}\psi _{\epsilon })
\end{equation}
holds. Then according to Remarks 1 and 2 the second term
of the last expression tends to zero as $\epsilon \to 0$,
therefore the statement of lemma B is proved.

{\bf The divergence of the sequence of regularized dynamics in the whole algebra $B(H)$}

The solution $u_{\epsilon}(t,x)$ of regularized problem  (\ref{S-Schr-2}) - (\ref{Cau-2b})
can be presented by the regulariszed group $U_{\epsilon}^t$ of
unitary operators in the Hilbert space ${\cal H}$:
$$u_{\epsilon}(t,x)=(U^t_{\epsilon})\varphi(x),$$ where
$U^t_{\epsilon}=e^{-it{\bf H}_{\epsilon}},\,t\in \mathbb{R}$ (see
(\ref{reggr})).

{\bf Theorem 2}. {If for some $t>0$ and $\varphi \in \cal H$ the sequence $\{ (U^t_{\epsilon})\varphi,\ {\epsilon \to +0}\}$ converges in weak topology of the space $\cal H$ but diverges in the strong topology then the sequence $\{(T^t_{\epsilon})\rho _{\varphi}\}$ diverges in topologe of point-wice convergence on the space $B(H)$. I.e. for any infinitesimal sequence $\{ \epsilon _k\}$ there is the operator ${\bf A}\in B(H)$ such that the numerical sequence $\{ \langle (T^t_{\epsilon})\rho _{\varphi},{\bf A}\rangle \}$ diverges.}

The proof of the theorem 2 follows from a more general theorem which is published in the paper \cite{SR11} (see theorem 14.2) or in the paper \cite{SFPM} (see theorem 7). Similar results is considered in the work \cite{D-A}.

{\bf The convergence of the sequence  regularized dynamics in the
algebra ${\cal A}_{comp}$}

Now we consider the dynamics of the states in the algebra ${\cal A}_{comp}$ of compact operators with the unity operator.

Let symbols $\Sigma ( H)$ and
$\Sigma _p(H)$ denote the set of state the set of pure states respectively on the algebra $B( H)$ of all bounded linear operators. Analogously, let  symbols $\Sigma ({\cal A})$ and
$\Sigma _p({\cal A})$ note the set of state the set of pure states respectively on some subalgebra $\cal A$ of algebra $B(H)$. Let symbol $\rho _u,\, u\in H,\, \|u\|=1,$ note the state on the algebra $\cal A$ acting by the equality $\langle \rho _u,{\bf A}\rangle =(u,{\bf A}u),\, {\bf A}\in B(H)$.

Let us define by means of degenerate Hamiltonian ${\bf H}=ibd/dx$ the following one-parameter family
$T^t_{comp},\, t\in \mathbb{R},$ of maps of the subset $\Sigma ({\cal
A}_{comp})$ into itself.

1. Case $b\leq 0$.

If $t\geq 0$ then $T_{comp}^t\rho _{\varphi }=\rho _{e^{-it{\bf
H}}\varphi }$ where $\rho _{e^{-it{\bf H}}\varphi }({\bf
A})=(e^{-it{\bf H}}\varphi ,{\bf A}e^{-it{\bf H}}\varphi )$ for
any ${\bf A}\in \cal A, \varphi\in H$.

If $t\leq 0$ then $T_{comp}^tJ=J$ and $T_{comp}^t\rho _{\varphi }=\rho _{e^{-it{\bf
H}^*}\varphi }+(1-\| e^{-it{\bf H}^*}\varphi \|^2)J$ where $\rho
_{e^{-it{\bf H}^*}\varphi }({\bf A})=(e^{-it{\bf H}^*}\varphi
,{\bf A}e^{-it{\bf H}^*}\varphi )$ for any ${\bf A}\in {\cal
A}_{comp}$ and $J({\bf I})=1;\ J({\bf A})=0$ for all compact
operators $\bf A$.

1. Case $b\geq 0$.

If $t\leq 0$ then $T_{comp}^t\rho _{\varphi }=\rho _{e^{-it{\bf
H}}\varphi }$ where $\rho _{e^{-it{\bf H}}\varphi }({\bf
A})=(e^{-it{\bf H}}\varphi ,{\bf A}e^{-it{\bf H}}\varphi )$ for
any ${\bf A}\in \cal A$.

If $t\geq 0$ then $T_{comp}^tJ=J$ and $T_{comp}^t\rho _{\varphi }=\rho _{e^{-it{\bf
H}^*}\varphi }+(1-\| e^{-it{\bf H}^*}\varphi \|^2)J$ where $\rho
_{e^{-it{\bf H}^*}\varphi }({\bf A})=(e^{-it{\bf H}^*}\varphi
,{\bf A}e^{-it{\bf H}^*}\varphi )$ for any ${\bf A}\in \cal A$.

Let for any $\epsilon \in (0,1)$ the regularized group  $T_{\epsilon}^t,\, t\in \mathbb{R},$ of transformation of the set $\Sigma _p({\cal H})$ is generated by the regularized Hamiltonian ${\bf L}_{\epsilon }$ by
the equality $T_{\epsilon}^t\rho _{\varphi
}=\rho_{\varphi,\epsilon}^t$ where
\begin{equation}\label{States-exp-3}
\rho_{\varphi,\epsilon}^t({\bf A})= <U^t_{\epsilon}\varphi,{\bf
A}U^t_{\epsilon}\varphi>\equiv \langle T_{\epsilon}^t\rho
_{\varphi }, {\bf A}\rangle ,\ \forall \ {\bf A}\in B(H),
\end{equation}
and $U^t_{\epsilon}=e^{-it{\bf H}_{\epsilon}},\, t\in R$ (see
((\ref{reggr}))).

Since the one-parameter families of maps are defined on the set of the states $\{ \rho _{\varphi },\, \varphi \in {\cal H}$ and this set of states is dense in the set $\Sigma ({\cal A}_{comp})$ in the topology of point-wice convergence (see \cite{D-A, SR11}) then it can be uniquely extended onto the whole set $\Sigma ({\cal A}_{comp}).$

{\bf Theorem 3}. {\it
The sequence of
regularized groups $\{ T_{\epsilon}^t\}$ converges to the one-parameter family of maps $T_{comp}^t$ in the following sense:

for any ${\bf A}\in {\cal A}_{comp}$, (of the type
 ${\bf A}=\lambda \cdot 1+K$) the equality
\begin{equation}\label{Limit-exp-3}
\lim_{\epsilon\to 0} \sup\limits_{t\in [-L,L]}| \langle
T_{\epsilon}^t\rho_{\varphi},{\bf A}\rangle - \langle
T_{comp}^t\rho _{\varphi }, {\bf A}\rangle|=0.
\end{equation}
holds for any $L>0, \varphi\in H$.
}

The statement of the theorem 3 is the consequence of the lemma 5 and theorem 1.

{\bf Proof}. Let ${\bf K}$ be a compact self-adjoint operator in
the space $H$ and let $\sigma $ is a positive number. Then there
is the finite number of unique vectors $\phi _1,...,\phi _m$ and
the finite number of real numbers $c_1,...,c_m$ such that $\|{\bf
K}-\sum\limits_{k=1}^mc_k{\bf P}_{\phi _k}\|_{B(H)}<{1\over
2}\sigma $ where ${\bf P}_{\phi _k}$ is the operator of orthogonal
projection on the one-dimensional subspace ${\rm lin}(\phi _k)$ of
the space $\cal H$.

Therefore $|\rho ^t_{\varphi ,\epsilon}({\bf K}) -\rho ^t_{\varphi ,\epsilon}(\sum\limits_{k=1}^mc_k{\bf P}_{\phi _k})|\leq {1\over 2}\sigma $. Since $\rho ^t_{\varphi ,\epsilon}(\sum\limits_{k=1}^mc_k{\bf P}_{\phi _k})=\sum\limits_{k=1}^mc_k\rho ^t_{\varphi ,\epsilon}({\bf P}_{\phi _k})=\sum\limits_{k=1}^mc_k|(u_{\epsilon }(t),\phi _k)|^2$ then according to the theorem 1 and Rieman oscilation theorem  $\lim\limits_{\epsilon \to 0}\rho ^t_{\varphi ,\epsilon}(\sum\limits_{k=1}^mc_k{\bf P}_{\phi _k})=\sum\limits_{k=1}^mc_k|\lim\limits_{\epsilon \to 0}[(u_{\epsilon }(t),\phi _k)]|^2=\sum\limits_{k=1}^mc_k|\int\limits_0^{+\infty }(\varphi (x+bt)\phi _k(x))dx|^2$.

If $u^*(t,x)=\phi (x+bt),\, (t,x)\in \mathbb{R}_+\times \mathbb{R}_+$, then
$\lim\limits_{\epsilon \to 0}\rho ^t_{\varphi ,\epsilon}(\sum\limits_{k=1}^mc_k{\bf P}_{\phi _k})=(u^*(t,\cdot),(\sum\limits_{k=1}^mc_k{\bf P}_{\phi _k})u^*(t,\cdot))$.

Since $|(u^*(t,\cdot),{\bf K}u^*(t,\cdot))-(u^*(t,\cdot),(\sum\limits_{k=1}^mc_k{\bf P}_{\phi _k})u^*(t,\cdot))|<{1\over 2}\sigma $ then

${\overline {\lim\limits_{\epsilon \to 0} }}|(u^*(t,\cdot),{\bf K}u^*(t,\cdot))-\rho ^t_{\varphi ,\epsilon}({\bf K})|\leq \sigma $.

Since $\sigma >0$ is arbitrary number then $\lim\limits_{\epsilon \to 0}\rho ^t_{\varphi ,\epsilon}({\bf K})=(u^*(t,\cdot),{\bf K}u^*(t,\cdot))$ for arbitrary $t\geq 0$ and arbitrary compact operator $\bf K$. The theorem 3 is proved.

{\bf Theorem 4}. {\it
The family of maps $T_{comp}^t,\, t\in R$, is not a group. But the restrictions $T_{comp}^t(b),\,
t\in R_{\pm}$ are both semigroups  of maps of the set $\Sigma
({\cal A}_{comp})$ into itself:
$T^t_{comp} T^{\tau}_{comp}=T^{t+\tau}_{comp}, \,t,\tau\in
\mathbb{R}_+.$
}

{\bf Proof}.
The function $T_{comp}^t,\, t\in \mathbb{R}$, is not a group since, in particular,  the
equalities $T_{comp}^{-t}T_{comp}^t\rho _{\varphi }= \rho
_{\varphi }= T_{comp}^tT_{comp}^{-t}\rho _{\varphi }$ can't
hold both for any $t>0$. The semigroup properties of the limit one-parameter family of maps $T_{comp}^t,\, t\in \mathbb{R}_{\pm },$ is the consequence of the semigroup properties of operator-functions $e^{-it{\bf H}^*},\, t\geq 0$ with $b < 0$, and $e^{-it{\bf H}},\, t\geq 0$ with $b>0$.

{\bf Remark 3.} The state  $\sigma_{\varphi}^t=T_{comp}^t\rho
_{\varphi }$  on the $C^*$-algebra ${\cal K}$ can be extended onto
the hole $C^*$-algebra $B(H)$.

{\bf Remark 4.} Only one of two semigroup $T_{comp}^t,\, t\in \mathbb{R}_{\pm },$ is the semigroup of inverse maps.

{\bf The convergence of the sequence of regularized state dynamics on the algebra ${\cal A}_{mult}$ and the failure of the semigroup properties}

{\bf Theorem 5}. {\it Let $b>0$ and let ${\cal A}_{mult}$ be the abelian $C^*$-algebra of operators ${\bf A}_f$ of  multiplication on the arbitrary measurable essentially bounded function $f(x)\in L_{\infty }(\mathbb{R}_+)$ in the space $L_2(\mathbb{R}_+)$. Then for any $\varphi \in L_2(\mathbb{R}_+)$ and for any operator ${\bf A}_f\in {\cal A}_{mult}$ the following equality
\begin{equation}\label{Limit-exp-6}
\lim\limits_{\epsilon \to 0}\rho _{\varphi, {\epsilon }}^t ({\bf A}_f) \equiv \langle T^t_{mult}\rho _{\varphi },{\bf A}_f\rangle =
\int_{0}^{\infty}f(x)|\varphi(bt+x)|^2dx+
\int_{0}^{bt}f(x)|\varphi(bt-x)|^2dx.
\end{equation}
holds for any $t\geq 0$.}

{\bf Theorem 6}. {\it
The one-parameter family of maps $T^t_{mult},\, t\geq 0,$ acting in the set $\Sigma ({\cal A}_{mult})$ according to (\ref{Limit-exp-6}), does not satisfy the semigroup property
$T^t_{mult} T^{\tau}_{mult}=T^{t+\tau}_{mult}, \,t,\tau\in \mathbb{R}_+.$

The dynamical maps $T^t_{mult},\, t\geq 0,$ can be presented by two one-parameter family of maps in the space $H=L_2(\mathbb{R}_+)$:

1) the first   one-parameter family of maps ${\bf V}_t,\, t\geq 0,$
acting in the space $H=L_2(\mathbb{R}_+)$ by the formula

${\bf V}_tu(x)=u(x+bt),\, x\geq 0$;

2) the second  one-parameter family of maps ${\bf W}_t,\, t\geq 0,$
acting in the space $H=L_2(\mathbb{R}_+)$ by the formular

${\bf W}_t\varphi(x)=\varphi(bt-x),\, x\in [0,bt],$

${\bf W}_t\varphi(x)=0,\, x> bt.$ }

{\bf Proof}. Let us note that ${\bf V}_t^*{\bf V}_t={\bf P}_{[bt,+\infty )}$ anf ${\bf W}_t^*={\bf W}_t,\, forall \, t\geq 0,$ ${\bf W}_0=\bf 0$ and  ${\bf W}_t^*{\bf W}_t={\bf P}_{[0,bt]}$.
Then, in particular
\begin{equation}\label{2-relat-0}
{\bf W}^t{\bf W}^{0}={\bf W}^{0}{\bf W}^t={\bf 0}\neq  {\bf W}^t;\quad {\bf W}^t{\bf W}^{t}={\bf P}_{[0,bt])}\neq {\bf W}^{2t },  \,\,\,t\in
\mathbb{R}_+.
\end{equation}

\medskip

{\bf Theorem 7}. {\it
Let $b>0$ and let ${\cal A}_{mult}$ be
the abelian $C^*$-algebra of operators ${\bf A}_f$ of
multiplication on the measurable essentially bounded function
$f(x)\in L_{\infty }(R_+)$ in the space $L_2(R_+)$. Then the equality
\begin{equation}\label{Kraus}
T^t_{mult} \rho ={\bf V}_t\rho {\bf V}_t^*+{\bf W}_t\rho {\bf W}_t^*\quad \forall \ t\geq 0,\, \rho \in \Sigma ({\cal A}_{mult}).
\end{equation}
holds.
Here
${\bf V}_t^*{\bf V}_t+{\bf W}_t^*{\bf W}_t={\bf I}$.
}

The equality (\ref{Limit-exp-6}) is the consequence of Lemma 5
and the expression of the functions $\psi _{\epsilon}$. In fact,
according to (\ref{lemma2}) and Remarks 1 and 2 we have

$$ (u_{\epsilon},{\bf A}u_{\epsilon })=(\phi_{\epsilon},{\bf A}\phi
_{\epsilon })+2{\rm Re}((\psi _{\epsilon},{\bf A}\phi _{\epsilon
}))+(\psi _{\epsilon},{\bf A}\psi _{\epsilon })=$$
$$=(v,{\bf A}v)+2{\rm
Re}(v,{\bf A}r _{\epsilon })+(r_{\epsilon},{\bf A}r _{\epsilon
})+2{\rm Re}((\psi _{\epsilon},{\bf A}v))+2{\rm Re}((\psi
_{\epsilon},{\bf A}r _{\epsilon }))+(\psi _{\epsilon},{\bf A}\psi
_{\epsilon }), $$ where $r_{\epsilon}=\phi _{\epsilon}-v$.

Since
$\| r_e\|_H\to 0$ and $\lim\limits_{\epsilon \to 0}(\psi
_{\epsilon},g)=0$
for any vector $g\in H$ then the equality
(\ref{Limit-exp-6}) holds. Then for any
$\varphi \in L_2(R_+)$ and for any operator ${\bf A}_f\in {\cal
A}_{mult}$ the following equality
\begin{equation}\label{Limit-exp-?}
\lim\limits_{\epsilon \to 0}\rho _{\varphi, {\epsilon }}^t ({\bf
A}_f) = ({\bf V}_t\varphi ,{\bf A}_f{\bf V}_t\varphi )+({\bf
W}_t\varphi ,{\bf A}_f{\bf W}_t\varphi )
\end{equation}
holds. Hence the theorem 7 and the equality (\ref{Kraus}) is proved.

The absence of semigroup property $T^t T^{\tau}\neq T^{t+\tau}, \,t,\tau\in \mathbb{R}_+$  is the consequence of the following equalities: if $T^t\phi(x)=\phi(bt-x)$ for any $\phi \in H$ with ${\rm supp}\phi \in [0,bt]$ then
 $T^{\tau } \phi (bt-x)=\phi(b\tau -bt+x)\neq \phi (b(\tau +t)-x)$.
The equality (\ref{2-relat-0}) for the function ${\bf W}^t,\, t\geq
0,$ can be easy checked.

\medskip

{\bf Remark 5}. The dynamics (\ref{Limit-exp-?}) of the set of states $\Sigma ({\cal A}_{mult})$ on the obsevable algebra ${\cal A}_{mult}$ is the realization of Kraus representation of one-parametric family of completely positive mapping of the set $\Sigma ({\cal A}_{mult})$, i.e. the representation of the type
\begin{equation}\label{42}
T^t\rho =\sum\limits_j{\bf V}_j(t)\rho {\bf V}_j(t)^*,
\end{equation}
where $\{{\bf V}_j(t)\}$ are bounded linear operators satisfying the condition
\begin{equation}\label{43}
\sum\limits_j{\bf V}_j(t)^* {\bf V}_j(t) ={\bf I}.
\end{equation}
Since the operators ${\bf V}_t$ and ${\bf W}_t$ in the formula (\ref{Limit-exp-?}) satisfy the equalities ${\bf V}^*_t{\bf V}_t={\bf P}_{[bt,+\infty )}={\bf P}_{H_0}(t)$ and ${\bf  W}_t^*{\bf W}_t={\bf P}_{[0,bt )}={\bf P}_{H_1}(t)$, then according to Wold (see \cite{SNF}) decomposition  $H=H_0(t)\oplus H_1(t)$ for isometric operators $\exp (it{\bf L})$ the pair of operators ${\bf V}_b(t),\, {\bf W}(t)$ satisfies the condition (\ref{43}). Hence the presentation (\ref{Limit-exp-?}) of the limit dynamics on $\Sigma ({\cal A}_{mult})$ is the realization of Kraus decomposition (\ref{42}) with two terms.
It should be noted that the first family of operators ${\bf V}_b(t),\, t\geq 0,$ is one-parametric $C_0$-semigroup but the second family of operators ${\bf W}(t),\, t\geq 0,$  is not satisfy the semigroup property according to the relations (39), (40). Thus we obtain that the limit dynamic $T^t,\, t\geq 0,$ of the set of states on the algebra ${\cal A}_{mult}$ is the Kraus decomposition of one-parameter family of completely positive mappings of the set $\Sigma ({\cal A}_{mult})$ with two terms.

\bigskip

{\bf The pure and mixed states on the algebras  ${\cal A}_{comp}$}

Any element $\bf A$ of the ring ${\cal K}_{comp}$ of compact self-adjoint operators is defined by the ortho normal basis $\{ e_k\}$ of eigenvectors of compact self-adjoint  operator  $\bf A$ and the sequence of its eigenvalues $\{ a_k\}\in c_0$, where $c_0$ is Banach space of infinitezimal sequences endowing with supremum-norm. Since $c_0^*=l_1$ then the space of linear functionals on the ring ${\cal K}_{comp}$ is the space of trace-class operators $T_1({\cal H})$. Any element   $\rho \in T_1({\cal H})$ is defined by the ortho normal basis of   eigenvectors of thace-class  self-adjoint  operator $\rho $ and  the sequence of its eigenvalues  $\{ p_k\}\in l_1$.

The functional $\rho$ on the space of compact operators $\sigma (H)$ is called continuous if for any sequence $\{ {\bf A}_n\}$ of compact operators the relation  $\rho ({\bf A}_n)\to 0$ is the consequence of the condition $\|{\bf A}_n\|_{B(H)}\to 0$. The functional $\rho$ on the space of compact operators  $\sigma (H)$ is called ultraweak continuous if the condition  $\rho ({\bf A}_n)\to \rho ({\bf A})$ satisfies for any sequence $\{ {\bf A}_n\}$ of compact operators which is monotone increasing to the operator ${\bf A}\in B(H)$ such that ${\bf A}_nx\to {\bf A}x\, \forall \, x\in {\cal H}$.

Let $\Sigma ({\cal K}_{comp})$ be the set of nonnegative continuuois functionals  $\rho$ on the ring ${\cal K}_{comp}$ such that $\rho ({\bf P}_n)\to 1$ for arbitrary sequence of projectors which monotone nondecreasing to unite operator.

{\bf Lemma 6}. {\it  The set  $\Sigma ({\cal K}_{comp})$ is the intersection of the unique sphere with the positive cone of the space of trace class operators  $T_1(H).$}

In fact any nonnegative element $\rho \in T_1(H)$ defines the nonnegative linear continuous functional on the ring ${\cal K}_{comp}$. Conversely any nonnegative linear continuous functional $\rho $ on the ring ${\cal K}_{comp}$ defines nonnegative bounded quadratic form on the space $\cal H$ such that associated with nonnegative self-adjoint trace class operator ${\bf A}_{\rho }\in T_1(H)$. Therefore the set $S ({\cal K}_{comp})$ of nonnegative linear continuous functionals on the ring ${\cal K}_{comp}$ coincides with the cone of nonnegative elements in the space $T_1(H)$.
The normalization condition for the functional $\rho \in S ({\cal K}_{comp})$ is equivalent to the belonging of nonnegative trace class operator to the unite sphere of the space  $T_1({\cal H})$.

Note that any continuous nonnegative functional $\rho \in S ({\cal K}_{comp})$ is ultraweak continuous.

Since the set $\Sigma ({\cal K}_{comp})$ of the states on the ring $\cal K$ is the convex compact in Banach space $T_1({\cal H})$ then the set of extreme points of the set $\Sigma ({\cal K}_{comp})$ coincides with the set of one dimentional orthogonal projectors.

\medskip

The algebra ${\cal A}_{comp}$ is the result of the adjointing of the unite operator  $\bf I$ to the  ring ${\cal K}_{comp}$. The set $\Sigma ({\cal K}_{comp})$ of the states on the ring ${\cal K}_{comp}$ is the part of the set of the states on the algebra ${\cal A}_{comp}$ and is the set of all ulraweak continuous states on the algebra ${\cal A}_{comp}$. The arbitrary state on the algebra   ${\cal A}_{comp}$ wich has no ulraweak continuity component is the state $J\in  \Sigma ({\cal A}_{comp})$ such that $J({\bf A})=0 \ \forall \ {\bf A}\in {\cal K}(H)$ and $J({\bf I})=1$.

{\bf Remark 6}. The state $J$ on the algebra of operators ${\cal A}_{comp}$ is not normal (i.e. is not ultraweak continuous) since for any sequence of finite dimentional projectors  $\{ {\bf P}_n\}$ which monotone increase to the unique operator $\bf I$ the condition ${\bf J} ({\bf P}_n)\to {\bf J}({\bf I})$ for $n\to \infty $ is not satisfied: ${\bf J} ({\bf P}_n)=0\ \forall \ n\in \bf N$ but ${\bf J}({\bf I})=1$ by the definition of the state $J$.

{\bf Lemma 7.} For any state $\rho \in  \Sigma ({\cal A}_{comp})$ there is the number  $\alpha \in [0,1]$ and the state $r\in  \Sigma ({\cal K}_{comp})$ such that $\rho =\alpha r+(1-\alpha )J$.

The proof of such statement can be finded in the paper
 \cite{PV}.

\bigskip

{\bf The destruction of pure states on algebras ${\cal A}_{comp}$ by the limiting dynamics}

{\bf Theorem 8.} Let $b>0$ and $T^t_{comp},\, t\geq 0,$, be the limit one-parametric semigroup of the maps of the set of states $\Sigma ({\cal A}_{comp})$ which is given by the equality (33) (see theorem 3). Then for any unique vector  $\varphi \in H$ there is the value  $$T_*=\sup \{ T>0:\ \| e^{-iT{\bf H}^*}\varphi \|_H=1\}$$
such that for any $t>T_*$
$$
T^t_{comp}\rho _{\varphi }=\alpha (t)\rho _{\Phi (t,\varphi )}+(1-\alpha (t))J,
$$
where $\alpha (t)=\|e^{-it{\bf H}^*}\varphi \|_H^2>0,$ and $\Phi (t,\varphi )=(\alpha (t))^{-{1\over 2}}e^{-it{\bf H}^*}\varphi $.

The statement of the theorem 8 is the consequence of the theorem 3; it means that the limit dynamics  $T^t_{comp},\, t\geq 0,$ of the set of the states $\Sigma ({\cal A}_{comp})$ transformes the pure state $\rho _{\varphi}$ into the convex combination of two different pure states. In this case the number  $T_*=T_*(\varphi )$ is the moment of destruction of the pure state.

For the example under the consideration in the theorem 1 the all objects of the statment of the theorem D has the explicit describtion, namely, $\alpha (t)=\int\limits_{bt}^{+\infty }|\varphi (x)|^2dx,$ $1-\alpha (t)=\int\limits_{0}^{bt}|\varphi (x)|^2dx,$ $T_*=\sup\{ t\geq 0:\ \int\limits_0^{bt}|\varphi (x)|^2dx=0\}$, и $e^{-it{\bf H}^*}\varphi (x)=\varphi (x+bt),\, x\in \mathbb{R}_+,\, t\geq 0.$

\bigskip

\section {Conclusions}
In the paper we studied the simple model of
Schrodinger equation with the degenerate Hamiltonian. As the
measure of simplicity of degenerated Hamiltonian we can consider
such characteristic of deviation of degenerated Hamiltonian from
the set of self-adjoint operators as the deficient indexes.

The deficient indexes of the degenerated Hamiltonian $\bf H$ in
the considered example (6) is equal to (1,0) or (0,1) for
different sign of the coefficient $b$ in the expression (7).

The examples of degenerated Hamiltonian with arbitrary finite
deficient indexes $(n_-,n_+)$ can be given by the first order
differential operator in the Cauchy problems for transport-type
equation which is considered in the work by \cite{DPL}.

For example, if the function $b$ in the expression (6) is given as
the ${{\pi}\over 2}$-periodic continuation of the function $\cos
(x),\, x\in [0,{{\pi }\over 2})$, and if $H=L_2([0,{{k\pi }\over
2})$ for some $k\in \bf N$ then the differential operator (6) with
the domain $D({\bf H})=\oplus_{j=1}^k\dot W^1_2([{{(j-1)\pi}\over
2},{{j\pi }\over 2}))$ is the symmetric operator with the
deficient indexes $(k,0)$. In this case the dynamic generated by
the operator $\bf H$ is the direct summa  of the $k$ independent
dynamics in the spaces $H_j=L_2([{{(j-1)\pi}\over 2},{{j\pi }\over
2})),\ j=1,...,k$, any of each is unitary equivalent to the
dynamic generated by the operator (6).

Similar results can be obtained for the Hamiltonian with the
degeneration on the exterior of the ball $B_R$ (see \cite{SR11},
example 6.3).

Any isometric semigroup $\bf U$ admits the Wold decomposition (see
\cite{SNF}) $H=H_0\oplus H_1$ such that the subspaces $H_0,H_1$ are reduce the operators ${\bf U}(t)$ of the semigroup $\bf U$,
the restriction ${\bf U}|_{H_0}$ is the unitary semigroup and the restriction ${\bf U}|_{H_1}$ is one-sided shift. The unitary component of dynamics is well-known object and the shift component is isomorphic to the dynamics which is presented by the isometric semigroup
$e^{-it{\bf H}^*}$ in the example in the part 2 with the operator $\bf H$ (see (6)) with $b>0$. In this sence the example (6) describes the characteristic properties of dynamics generated by the arbitrary degenered symmetric Hamiltonian with nontrivial indexes of deficience.

Thus if the degenerate Hamiltonian $\bf H$ is maximal symmetric operator then the dynamics generated by Schrodinger equation with this Hamiltonian is isomorphic to one of the two dynamics:

-- isometric but not unitary (in the case $n_+=0$) as in the example with the condition $b\leq 0$;

-- dissipative and destroying the pure states (in the case $n_+>0$) as in the example with $b> 0$.

\bigskip

\section{Acknowledgments}
This work is supported by the Russian Science Foundation under grant
14-11-00687 and performed in the Steklov Mathematical Institute of Russian Academy of Sciences,
Moscow, Russia.


\begin{thebibliography}{99}

\bibitem{Fichera } G. Fichera, On a unified theory of boundary value problems
for elliptic-parabolic equations of second order// Boundary
problems in differential equations, The University of
Wisconsin Press, Madison. "--- 1960. "--- P. 97--120.




\bibitem{Oleinik-Radkevich } O.A. Oleinik, E.V. Radkevich. Second order equations with nonnegative characteristic form. (Russian)
Itogi Nauki, Ser. Mat., Mat. Anal. 1969, 7-252 (1971).


\bibitem{Freidlin-Ventcel'} A.D. Ventcel',
M. I. Freidlin,
Some problems concerning stability under small random perturbations.
Theory of Probability and its Applications, 1973, 17:2, 269--283.

\bibitem{Sakb-Vol} I.V. Volovich, V.Zh. Sakbaev Universal boundary value problem for equations of mathematical physics. Proceedings of the Steklov Institute of Mathematics
2014, Volume 285, Issue 1, pp 56-80.

\bibitem{Sakb-Smol} V.Zh. Sakbaev, O.G. Smolyanov.
Diffusion and Quantum Dynamics of Particles with Position-Dependent Mass, Doklady Mathematics, Vol. 86, No. 1, 2012, pp. 460-463.



\bibitem{STMF} V.Zh. Sakbaev, Averaging of quantum dynamical semigroups. Theor. and Math. Phys. 2010. Т. 164:3. С. 1215-1221.


    \bibitem{SFPM} { V.Zh. Sakbaev}, {\it On dynamics of quantum states generated by the Cauchy problem for the Schredinger equation with degeneration on the half-line}//
Journal of Mathematical Sciences, 2008, Volume 151, Issue 1, pp 2741-2753.


\bibitem{SR11} { V.Zh. Sakbaev,} {\it On the Cauchy problem for linear differential equation with degeneration and the averaging of iits regularization.}
// {
Sovrem. Mat. Fundam. Napravl.} {\small\bf 43} (2012) 3--174.


\bibitem{Gadella} M. Gadella, S. Kuru, J. Negro. Self-adjoint Hamiltonians with a
mass jump: General matching conditions, {\it Phys. Letters
A}  {\bf  362} (2007) 265--268.









\bibitem{GS} M. Gadella,  O.G. Smolyanov.  Feynman Formulas for Particles with Position-Dependent Mass, Doklady Math. 77 (1) (2007) 120-123.










\bibitem{Shafarevich} V. L. Chernyshev, A. A. Tolchennikov, A. I. Shafarevich. 	Behavior of Quasi-particles on Hybrid Spaces. Relations to the Geometry of Geodesics and to the Problems of Analytic Number Theory,
Regul. Chaotic Dyn., 21:5 (2016),  531 - 537



\bibitem{DPL} {\it Di Perna, P. Lions} Ordinary differential equation,
transport theory and Sobolev spaces. Invent. Math. 1989. {\bf V}.
98. P. 511-547.



\bibitem{OV} {\it M. Ohya, I.V. Volovich.} Mathematical foundations of quantum information and computation and its applications to nano- and bio-systems, Springer, Dordrecht, 2011.

\bibitem{ALV} L. Accardi, Yu.G. Lu, and I. Volovich, {\it Quantum theory and its
stochastic limit}, Springer, 2002.

\bibitem{TV} A. S. Trushechkin, I. V. Volovich, Perturbative treatment of inter-site couplings in the local description of open quantum networks, EPL, 113:3 (2016), 30005.

\bibitem{AVK} I. Ya. Aref'eva, I. V. Volovich, S. V. Kozyrev, Stochastic limit method and interference in quantum many-particle systems, Theoret. and Math. Phys., 183:3 (2015), 782 - 799.

\bibitem{VK} I. V. Volovich, S. V. Kozyrev, Manipulation of states of a degenerate quantum system, Proc. Steklov Inst. Math., 294 (2016), 241 - 251.

\bibitem{BR}
{\it O. Bratteli, D.W. Robinson}. {\it Operator algebras and
quantum statistical mechanics}, Springer-Verlag, (1979).


\bibitem{Takesaki} {\it M. Takesaki. } On the conjugate space of operator algebra. Tohiku Math. J. V. 10 (1958), P. 194-203.


\bibitem{Wils} {\it W.I.M. Wils.}  Stone-Cech compactification and representations of operator algebras, http://hdl.handle.net/2066/107571.

\bibitem{PV} V.Zh. Sakbaev, {\it  On the variational description of the trajectories
of averaging quantum dynamical maps}, P-adic numbers, ultrametric
analysis and appl. 2012. {\bf V. 4}, N 2. P. 120-134.




\bibitem{D-A} {\it G.F. Dell' Antonio.} On the limits of sequences of normal states. Comm. Pure Appl. Math. V. 20. P. 413-429. 1967.



\bibitem{SNF} B. Sz.-Nagy, C. Foias. Analyse harmonique des
operateurs de l'espace de Hilbert. 1967.

\end{thebibliography}
\end{document}